\documentclass[amsmath,amssymb,twocolumn,prl]{revtex4-1}

\usepackage{graphicx}               % Include figure files                   
\usepackage{bm}                     % bold math                             
\usepackage{hyperref}               % for hyper reference link                
\begin{document}                    % main document begin here

\title{Molecular dynamics simulations of the evaporation of 
        particle-laden droplets }

\author{Weikang Chen$^{1,3}$}
\email{wchen@ccny.cuny.edu}

\author{Joel Koplik$^{1,3}$}
\email{koplik@sci.ccny.cuny.edu}

\author{Ilona Kretzschmar$^2$}
\email{kretzschmar@ccny.cuny.edu}

\affiliation{
 Benjamin Levich Institute$^1$ and Departments of Chemical 
 Engineering$^2$ and Physics$^3$\\ 
 City College of the City University of New York, New York, NY 10031
}

\date{\today}

\begin{abstract}
% 132words 
We use molecular dynamics simulations to study the evaporation 
of particle-laden droplets on a heated surface. The droplets 
are composed of a Lennard-Jones fluid containing rigid particles 
which are spherical sections of an atomic lattice, and heating is 
controlled through the temperature of an atomistic substrate. We 
observe that sufficiently large (but still nano-sized) particle-laden 
drops exhibit contact line pinning, measure the outward fluid flow 
field which advects particle to the drop rim, and find that the structure 
of the resulting aggregate varies with inter-particle interactions.  
In addition, the profile of the evaporative fluid flux is measured 
with and without particles present, and is also found to be in qualitative 
agreement with earlier theory.  The compatibility of simple nanoscale 
calculations and micron-scale experiments indicates that molecular 
simulation may be used to predict aggregate structure in evaporative 
growth processes.  
\end{abstract}


\maketitle

%%%%%%%%%%%%%%%%%%%%%%%%%%%%%%%%%%%%%%%%%%%%%%%%%%%%%%%%%%%%%%%%%%%%%%%%%%%%%
% Introduction
%%%%%%%%%%%%%%%%%%%%%%%%%%%%%%%%%%%%%%%%%%%%%%%%%%%%%%%%%%%%%%%%%%%%%%%%%%%%%
The evaporation of a sessile droplet on a hot surface is a key problem in 
fluid mechanics, relevant both to theoretical issues in heat transfer and to
practical questions in materials processing. The evaporation of a 
{\em particle-laden} droplet raises the additional issue of the structure of 
the resulting solid aggregate, and, going further, offers the possibility of   
controlling this structure by means of anisotropic ({\em e.g.}, Janus) surface
properties \cite{Jiang}.  A familiar and paradigmatic example of this 
process occurs in coffee stains, where the residue of evaporated 
droplets takes the form of a ring-like deposit of grains at the rim. 
Experiments by
Deegan and collaborators \cite{Deegan} focused attention on this 
``coffee ring problem'' several years ago, and subsequent work 
\cite{Deegan1,Bigioni,Bodiguel,Bhardwaj}
established the ubiquity of the process, while numerous theoretical 
studies have addressed the dynamics 
\cite{Deegan2,Popov,Okuzono,Cazabat,Kumar}.
A complete understanding of the problem is not yet available however: 
experiments
cannot measure everything in a small, time-dependent, multiphase droplet,  
while most theoretical treatments require approximations to deal with an
evaporating particle-laden drop.

In this paper we use molecular dynamics (MD) simulations to simulate the 
evaporation of droplets containing colloidal particles, having either uniform 
or Janus-like surface properties. One goal is to test whether the
phenomena found in micron-sized particle systems persist down to nanometer
scales; in this way we hope to extend the size range in which controlled
aggregate structures may be produced by droplet evaporation.  A second goal
is to test the validity of some of the underpinnings of the theoretical
analyses used in the problem. Since MD simulations provide detailed
atomic-scale information, we can measure concentration, temperature and 
fluid flow, even during the rapid heterogeneous processes occurring in
evaporation.  The difficulties of applying uncertain constitutive relations 
are absent, although replaced to some degree by the problem of extracting a
robust signal from a relatively small sample in a fluctuating environment. 
More generally, 
our goal is to establish the ability of these relatively basic
simulations of moderate scale systems to predict phenomena occurring in
droplet evaporation and guide experimental investigations. 

%%%%%%%%%%%%%%%%%%%%%%%%%%%%%%%%%%%%%%%%%%%%%%%%%%%%%%%%%%%%%%%%%%%%%%%%%%%%%
% Basic simulation
%%%%%%%%%%%%%%%%%%%%%%%%%%%%%%%%%%%%%%%%%%%%%%%%%%%%%%%%%%%%%%%%%%%%%%%%%%%%%
The simulations use standard molecular dynamics (MD) techniques
\cite{md1, md2, md3} and generic interactions of Lennard-Jones form,
\begin{equation}\label{eq:lj}
 V(r_{ij})=4\epsilon \left[ \left(\frac{\sigma}{r_{ij}}\right)^{12}
                        -\left(\frac{\sigma}{r_{ij}}\right)^{6}\right]
\end{equation}
For simplicity we assume that all fluid and particle atoms have the same 
interaction potential, along with the same mass $m$ and approximate diameter 
$\sigma$.  The calculations are nondimensionalized using $\epsilon$, $\sigma$ 
and $m$ as energy, length and mass scales, respectively, and the resulting 
time scale is $\tau = \sigma(m/\epsilon)^{1/2}$.  Typical numerical values are 
$\sigma\sim 0.3$nm, $\tau\sim 2$ps and $\epsilon\sim 120 k_B$, where $k_B$ is 
Boltzmann's constant, and temperatures are measured in units of 
$\epsilon/k_B$.  The fluid atoms in liquid or vapor obey ordinary
Newtonian dynamics with the force arising from the interaction with other
atoms (within a cutoff radius of 2.5$\sigma$), using a predictor-corrector
method for the integration of the equations of motion.  The particles are 
spherical sections of an atomic fcc lattice containing all atoms within a
certain radius of a center; here the atomic density is 0.8, the radius is 2
and the particles contain 32 atoms. These are Janus particles, implemented
by making the atoms in only one hemisphere of a particle attractive to other 
particle atoms; all particle atoms attract the fluid and wall in the same
way. The particles
move as rigid bodies, where the net force and torque on each particle
is computed by summing the interatomic forces between its atoms and the
neighboring fluid atoms, and the motion is given by the Newton 
and Euler equations. Quaternion variables are used to describe the particle
orientations \cite{md1}.  The solid substrate is made of atoms coupled to
fcc lattice sites by a linear spring of stiffness 100$\epsilon/\sigma^2$.

\begin{figure} [h]
  \includegraphics[width=0.36\textwidth,height=0.25\textwidth]{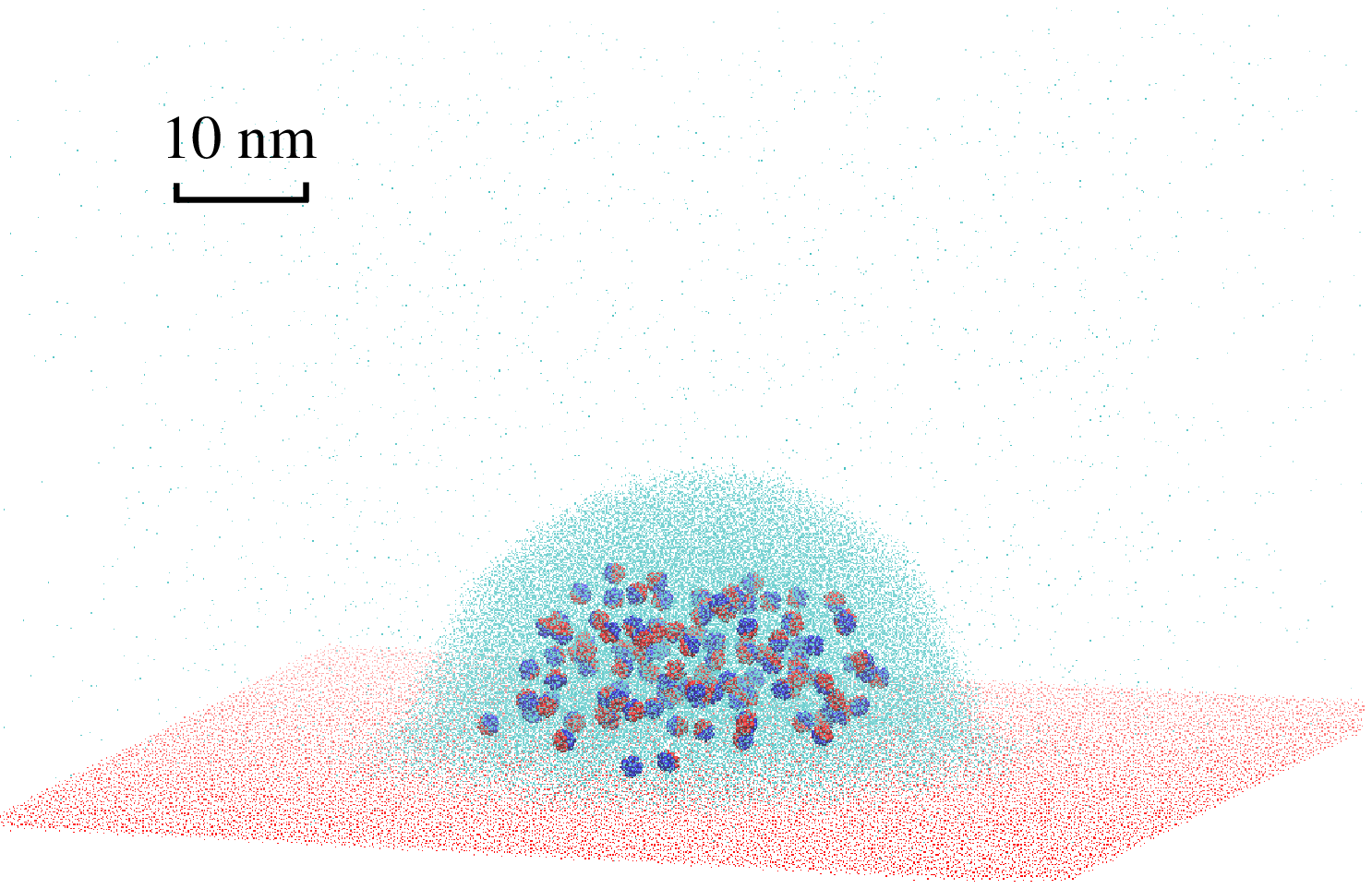}
  \includegraphics[width=0.36\textwidth,height=0.25\textwidth]{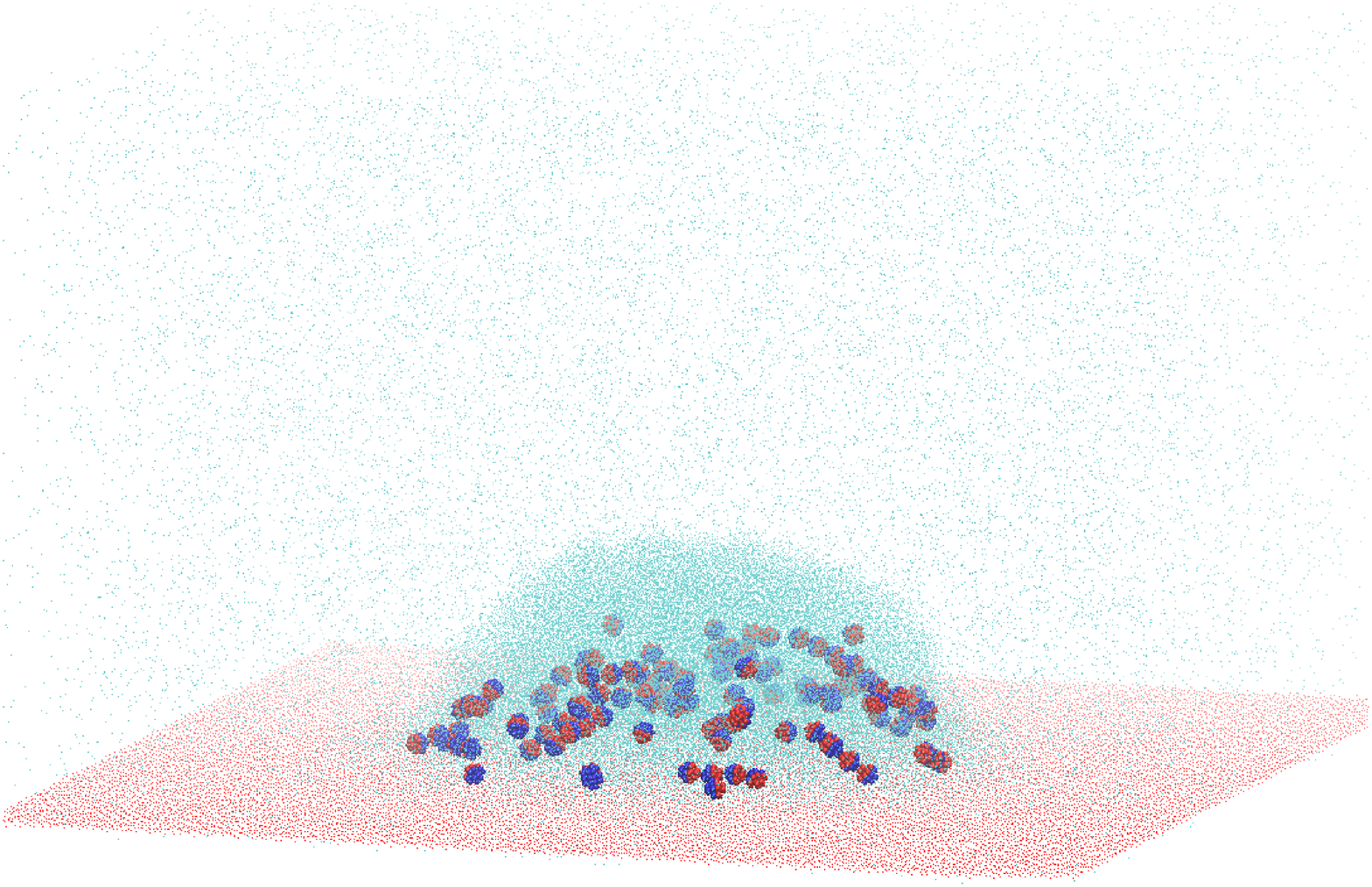}
  \includegraphics[width=0.36\textwidth,height=0.25\textwidth]{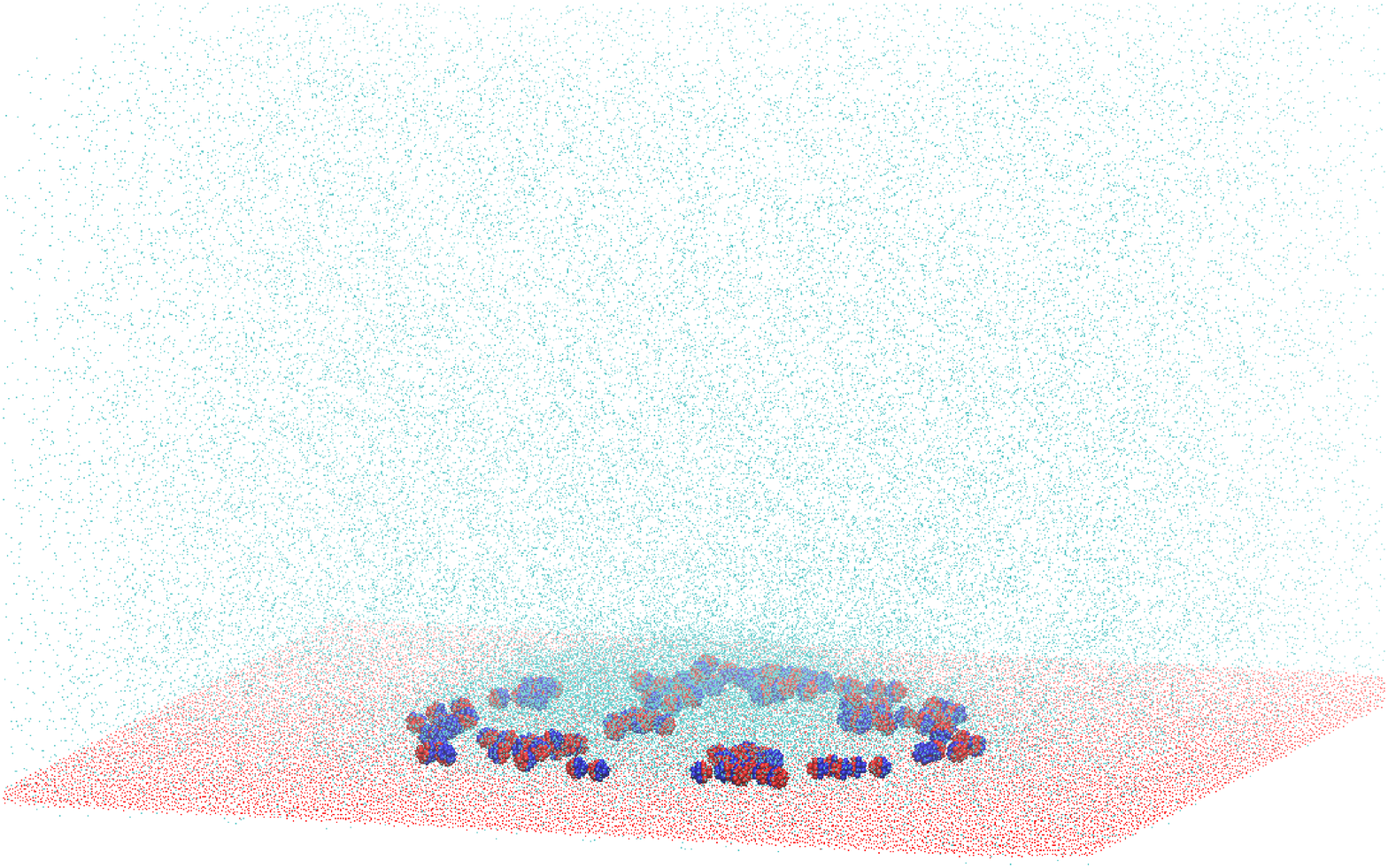}
  \caption{(Color online) Evaporation of a particle-laden droplet: fluid and
  solid atoms are shown as cyan and red dots respectively and the Janus 
  particles are circles whose two sides are red and blue. 
  Top to bottom: times 50, 500 and 1000$\tau$. \label{fig:evap}}
\end{figure}

\begin{figure} [h]
   \includegraphics[width=0.22\textwidth]{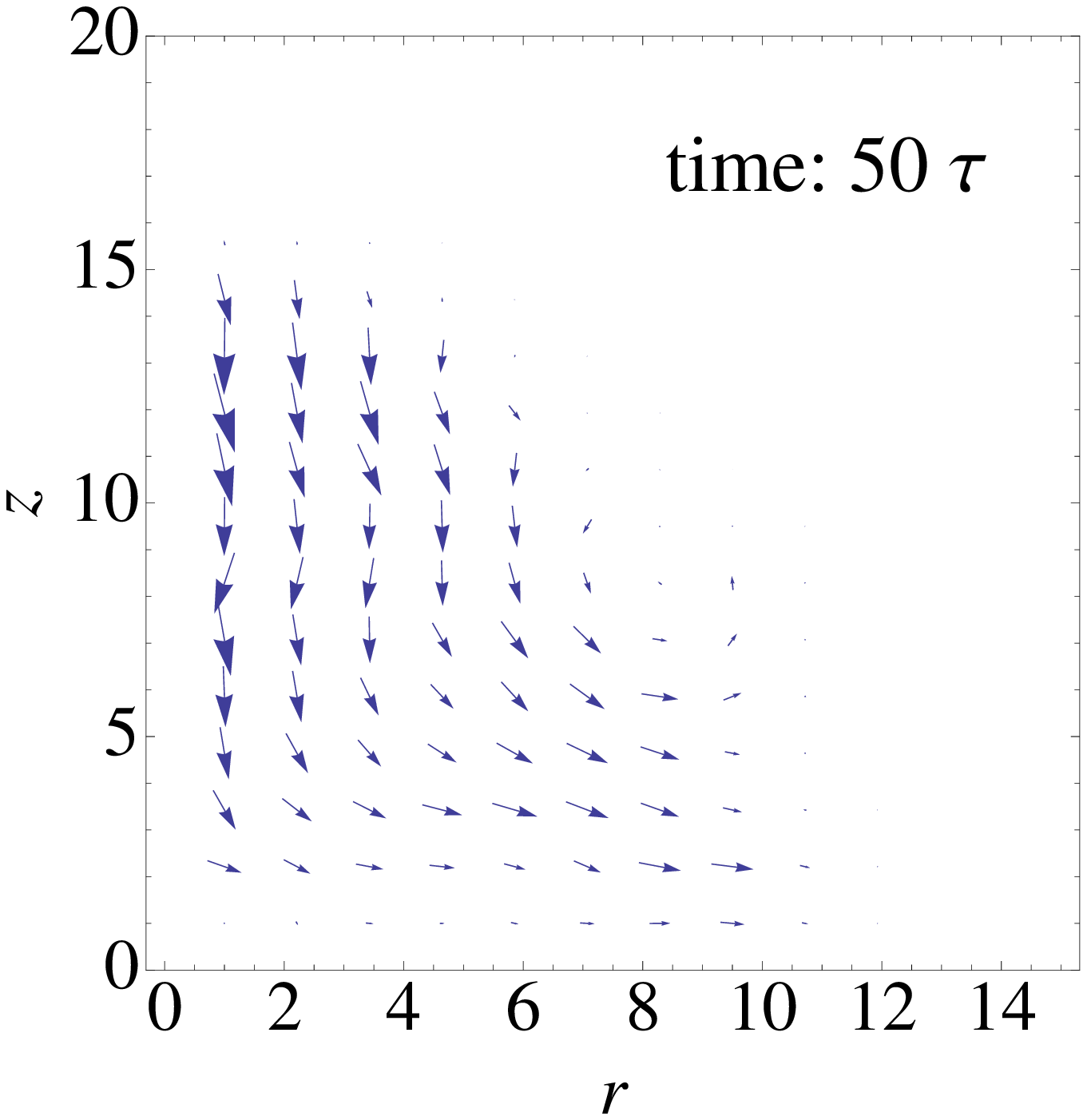}
   \includegraphics[width=0.22\textwidth]{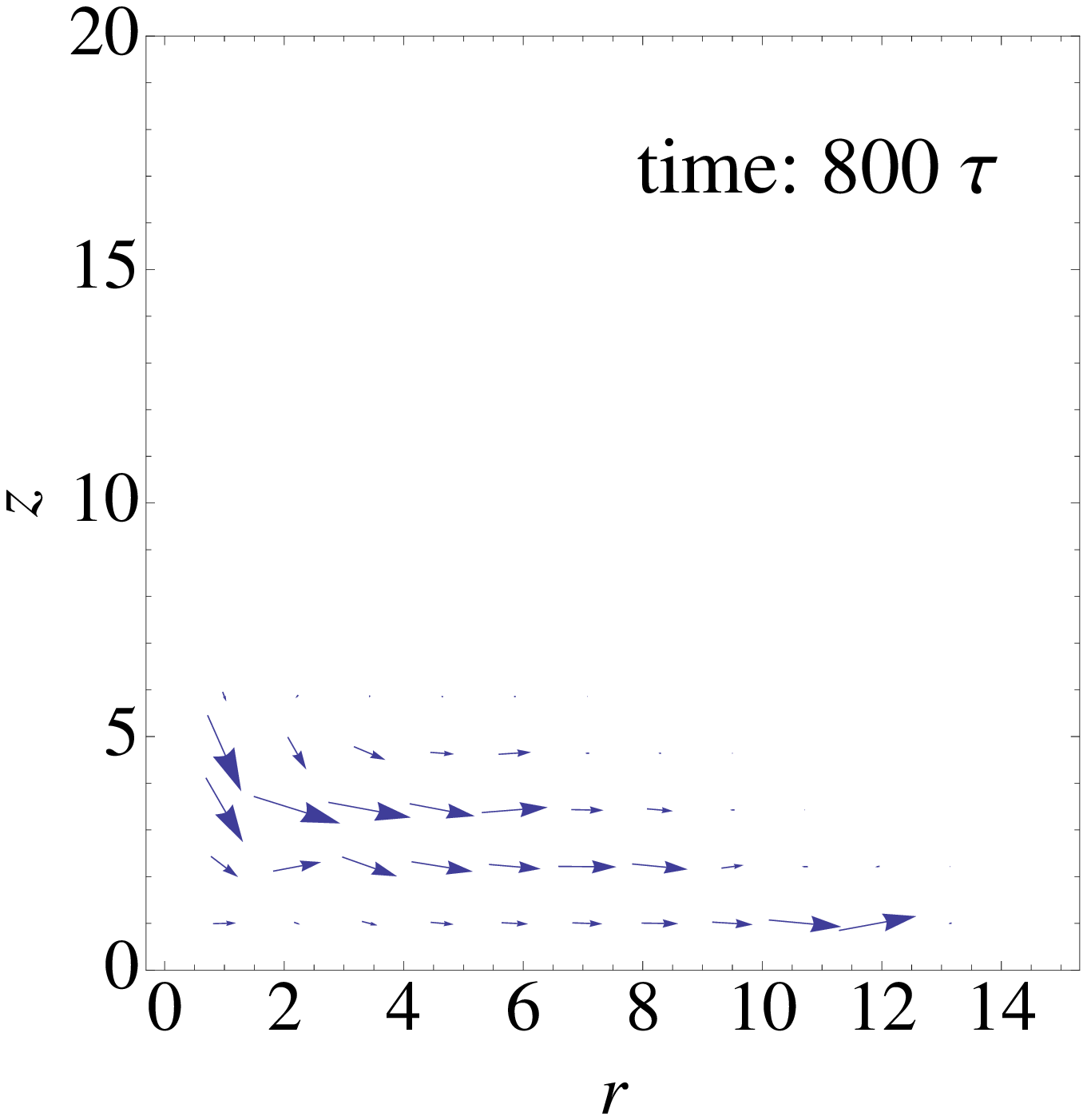}
   \caption{Velocity field in cylindrical coordinates in an evaporating 
   particle-laden droplet at early and late times; the solid occupies 
   the region $z<0$. \label{fig:velo}}
\end{figure}

Initially the drop consists of a hemispherical cap of 72,236 fluid atoms 
placed above a (monolayer) solid wall.  119 particles of 32 atoms each are 
centered at random positions within 
the cap, and the entire system is equilibrated by gradually raising the
temperature from 0.5 to 1.2 using velocity rescaling, following which the 
substrate is maintained at this temperature while the fluid temperature is
allowed to vary.  The liquid expands slightly during the temperature ramp
and then, as seen in Fig.~\ref{fig:evap}, the drop shrinks monotonically
as it emits vapor, and eventually disappears due to evaporation.  Evaporating 
fluid atoms which leave the simulation box are simply deleted.  Meanwhile the 
particles settle towards the substrate and subsequently are 
advected to the rim of the droplet and deposit there.  The contact line
itself remains pinned.  The connection
between liquid and particle motion is indicated by the velocity field 
shown in Fig.~\ref{fig:velo}: the fluid moves downward over most of the drop
and radially outward near the substrate.  This velocity field is computed by
dividing the simulation domain into concentric circular rings centered along
the vertical axis of the drop, and simply averaging the atomic velocities in
each ring.

%%%%%%%%%%%%%%%%%%%%%%%%%%%%%%%%%%%%%%%%%%%%%%%%%%%%%%%%%%%%%%%%%%%%%%%%%%%%%
% Evaporation
%%%%%%%%%%%%%%%%%%%%%%%%%%%%%%%%%%%%%%%%%%%%%%%%%%%%%%%%%%%%%%%%%%%%%%%%%%%%%

The origin of the flow that drives particles to the rim is, it is believed
\cite{Deegan}, contact line pinning coupled to the fact that the
evaporative flux is largest at the edges of the drop.  The liquid must
supply this flux as the droplet shrinks down, and the geometry of the situation 
requires a strong outward 
flow field, as seen in Fig.~\ref{fig:velo}.  More precisely, for
drops of pure liquid evaporating with a fixed contact line, one can show
\cite{Hu_JPC,Hu_Lang,Okuzono,Cazabat} that
\begin{equation} \label{eq:vp}
j(r, \theta)=j_0\left(1-\frac{r^2}{R^2}\right)^{-\lambda(\theta)}
\quad
\lambda(\theta)=\frac{1}{2}\left(1 - \frac{\theta}{\pi-\theta}\right)
\end{equation}
Here, $\theta$ is the contact angle between the droplet and the wall, which
varies as the droplet evaporates.  The key point is that for $\theta < \pi/2$
the (mathematical) vapor flux diverges at the edge of the droplet. 
Of course, there is no real singularity in a physical problem, and one
expects $j$ to be cut off at a small (molecular) scale.  

\begin{figure} [h]
  \begin{center}
    \includegraphics[width=0.42\textwidth]{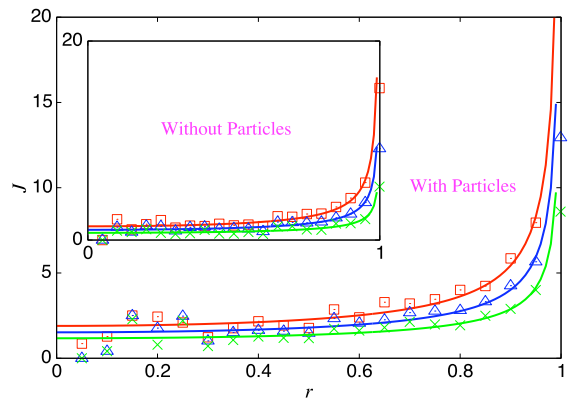}
    \caption{(Color online) Evaporative vapor flux vs. scaled radius 
    for particle laden-drops and, in the inset, for pure fluid drops. 
    The three curves refer to times 300$\tau$ ($\times$, green), 
    400$\tau$ ($\triangle$, blue) and 500$\tau$ ($\Box$, orange).  
    \label{fig:curr}}
  \end{center}
\end{figure}

We have measured the evaporative flux both for pure liquid and
particle-laden drops, and obtained 
Fig.~\ref{fig:curr}, which shows $j(r,t)$ at the three successive times
indicated.  Each plot is an average over a short (10$\tau$) interval
centered at 300, 400 and 500$\tau$. Averaging is necessary to smooth the
fluctuations, and longer averaging periods would be more effective, but 
the drop shape changes too much over longer intervals. The entire evaporation 
process lasts for about 1000$\tau$ (roughly 2 ns) for the Janus case, and 
slightly longer for pure fluid evaporation, but the data at later
times involves fewer and fewer evaporating atoms and is too noisy for analysis.
The ordinate in Fig.~\ref{fig:curr} is radial
position divided by the current drop radius, and varies between 0 and 1.
While the radius is constant (except for fluctuations) in the particle-laden
case, the radius of the pure fluid drop decreases with time.
We see that indeed the vapor current tends to diverge at the rim and the 
divergence increases with time. This behavior is qualitatively consistent
with Eq.~\ref{eq:vp}, since a stronger divergence requires a smaller contact
angle, and the simulations show this trend.  However, the formula fails to
quantitatively describe the simulations.  The exponents $\lambda(\theta)$
obtained in fitting the pure fluid data are 0.48, 0.55 and 0.60 at times 300, 
400 and 500$\tau$, respectively, but Eq.~\ref{eq:vp} does not permit values 
$\lambda>1/2$.  Some possible explanations for this discrepancy are
deviations from bulk continuum behavior in nanosized regions, difficulties 
in measuring the flux accurately in directions nearly parallel to the
surface, and inadequate ensemble averaging in obtaining the data.
The flux is slightly more singular for the particle-laden drop: the fitted 
values of $\lambda$ are 0.53, 0.58 and 0.62 for the same three times. 

\begin{figure} [h]
  \includegraphics[width=0.23\textwidth,height=0.23\textwidth]{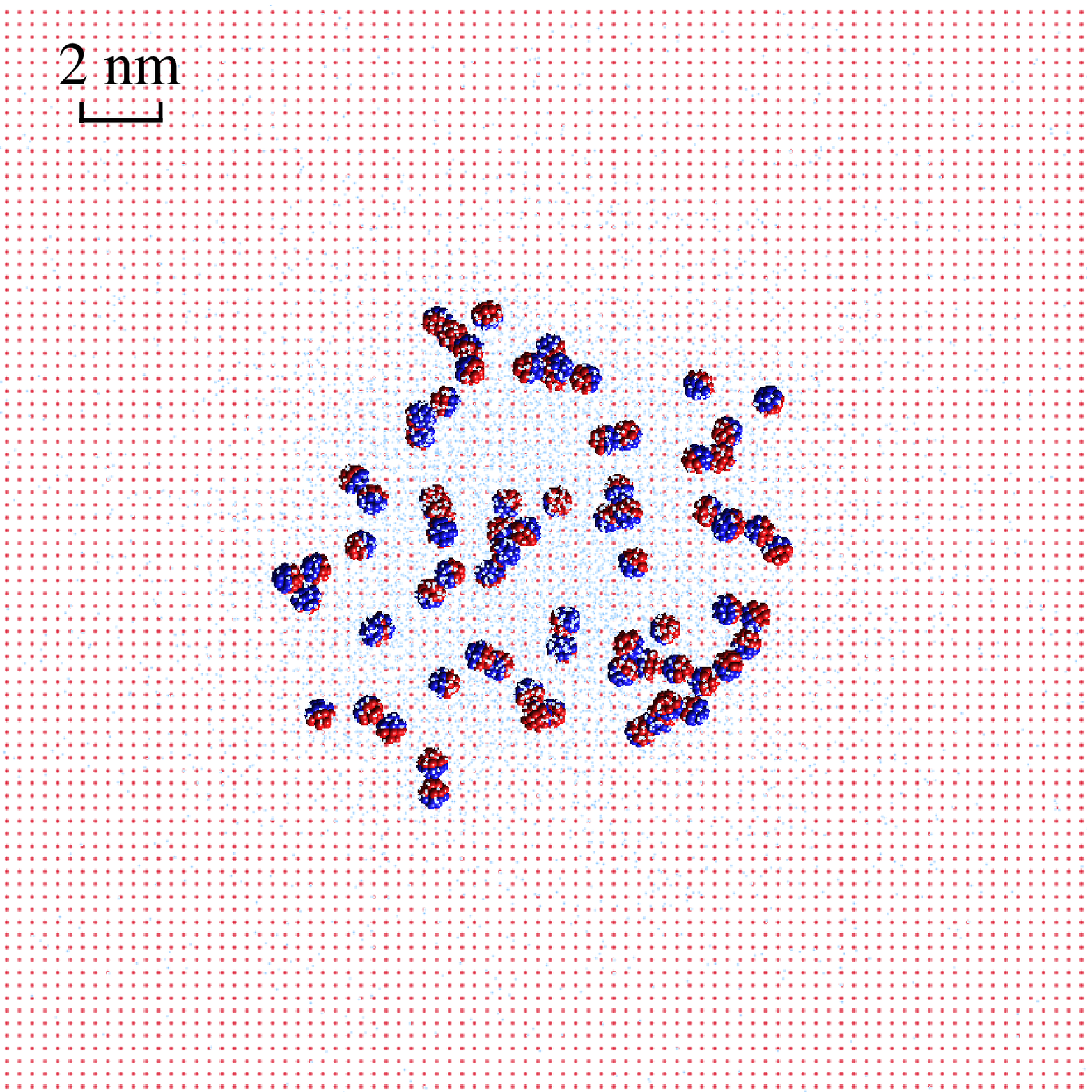}
  \includegraphics[width=0.23\textwidth,height=0.23\textwidth]{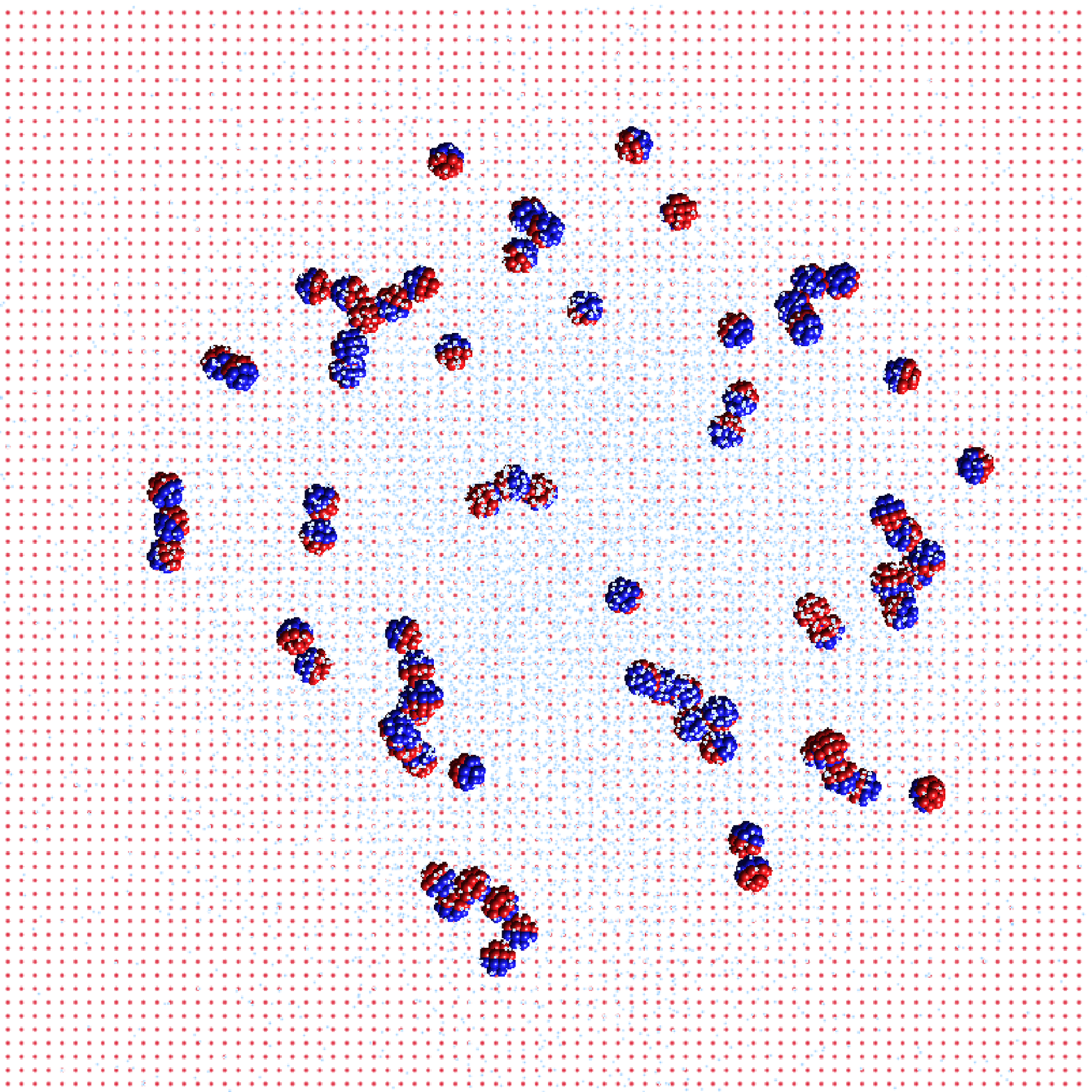}
  \caption{(Color online) Particle deposit for evaporated drops of initial 
  radius (left) 5 and (right) 10 nm. The color scheme is given in the
  caption to Fig.~\ref{fig:evap}. \label{fig:min}}
\end{figure}

%%%%%%%%%%%%%%%%%%%%%%%%%%%%%%%%%%%%%%%%%%%%%%%%%%%%%%%%%%%%%%%%%%%%%%%%%%%%%
% Variant cases
%%%%%%%%%%%%%%%%%%%%%%%%%%%%%%%%%%%%%%%%%%%%%%%%%%%%%%%%%%%%%%%%%%%%%%%%%%%%%
Not all evaporating nanodrops behave in this way.  One important requirement
for obtaining a deposit at the rim is that the drop be large enough:  if the
drop is too small it evaporates, or at least decays into a thin pancake,
before the flow field is established and the particles are able to move to the
rim.  This behavior was first observed experimentally by Shen {\em et al}. 
\cite{Shen}, and we have reproduced it in simulations.  The drop shown in
Fig.~\ref{fig:evap} has a radius of about 20 nm, whereas in similar 
simulations for drops whose initial radius is 5 or 10 nm we see that the
particles deposit roughly uniformly over the drops interior;  see
Fig.~\ref{fig:min}.

A second constraint needed for deposition to occur at the drop rim is that the
liquid must have adequate thermal contact with the solid to set up the flow 
field seen above \cite{Kim}.  We have investigated this issue by varying the 
interaction between the liquid and the wall:  in the Lennard Jones potential 
(Eq.~\ref{eq:lj}) acting between fluid and wall atoms, we varied the
coefficient of the attractive $r^{-6}$ term between 0 (pure short-distance 
repulsion - hydrophobic wall) and 1 (standard strength attraction -
completely wetting wall)  and observed the resulting solid pattern while
measuring the thermal conductivity.  The latter simulation involved a slab
of liquid completely filling the gap between two atomic walls held at
different temperatures.  As the attractive strength decreased from 1 to 0
the thermal conductivity decreased approximately linearly, ultimately
by a factor of nearly 10.  Correspondingly, as the wall attractive strength 
decreased the resulting pattern of 
solid particles varied from the rim deposit shown above to a
crystallized droplet. The effects of substrate thermal resistance and 
conductivity on evaporation have been studied more systematically by 
Dunn {\em et al}. \cite{Dunn1,Dunn2}. 

%%%%%%%%%%%%%%%%%%%%%%%%%%%%%%%%%%%%%%%%%%%%%%%%%%%%%%%%%%%%%%%%%%%%%%%%%%%%%
%Deposition pattern
%%%%%%%%%%%%%%%%%%%%%%%%%%%%%%%%%%%%%%%%%%%%%%%%%%%%%%%%%%%%%%%%%%%%%%%%%%%%%

The detailed structure of the deposit is an important consideration in
potential applications to evaporative self-assembly. We saw above that
particles with Janus surface properties would, under the right conditions,
form a rim deposit with chain-like structures.  This behavior 
is confirmed by experimental observations of gold-capped sulfated-polystyrene 
Janus particle-laden droplets during drying \cite{Guzman}. 
The chaining behavior is a result of the attractive interaction between the 
gold caps of and the sulfated polystyrene half of the Janus particles.

\begin{figure} [h]
  \includegraphics[width=0.30\textwidth, height=0.30\textwidth]{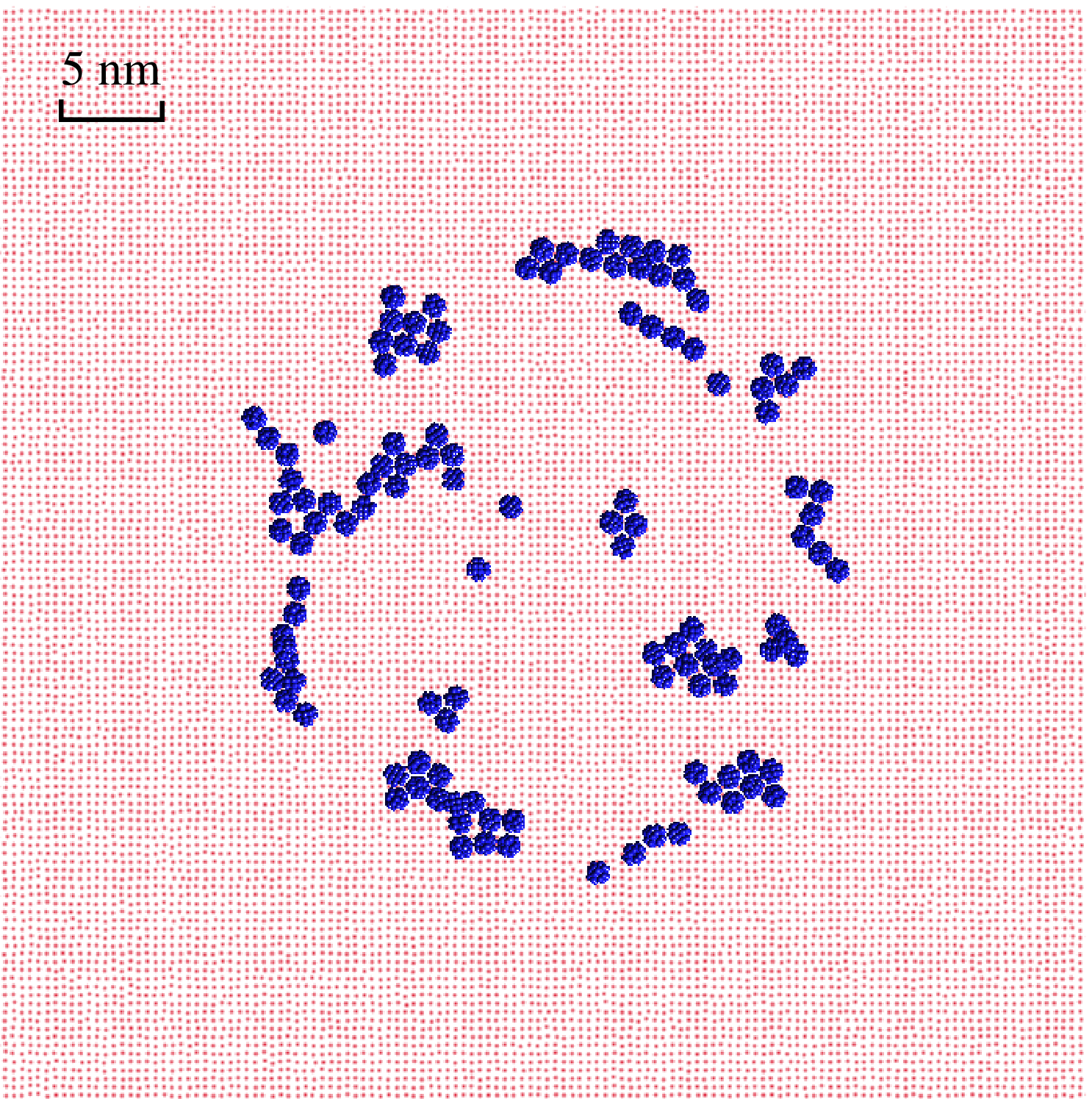} 
  \includegraphics[width=0.30\textwidth, height=0.30\textwidth]{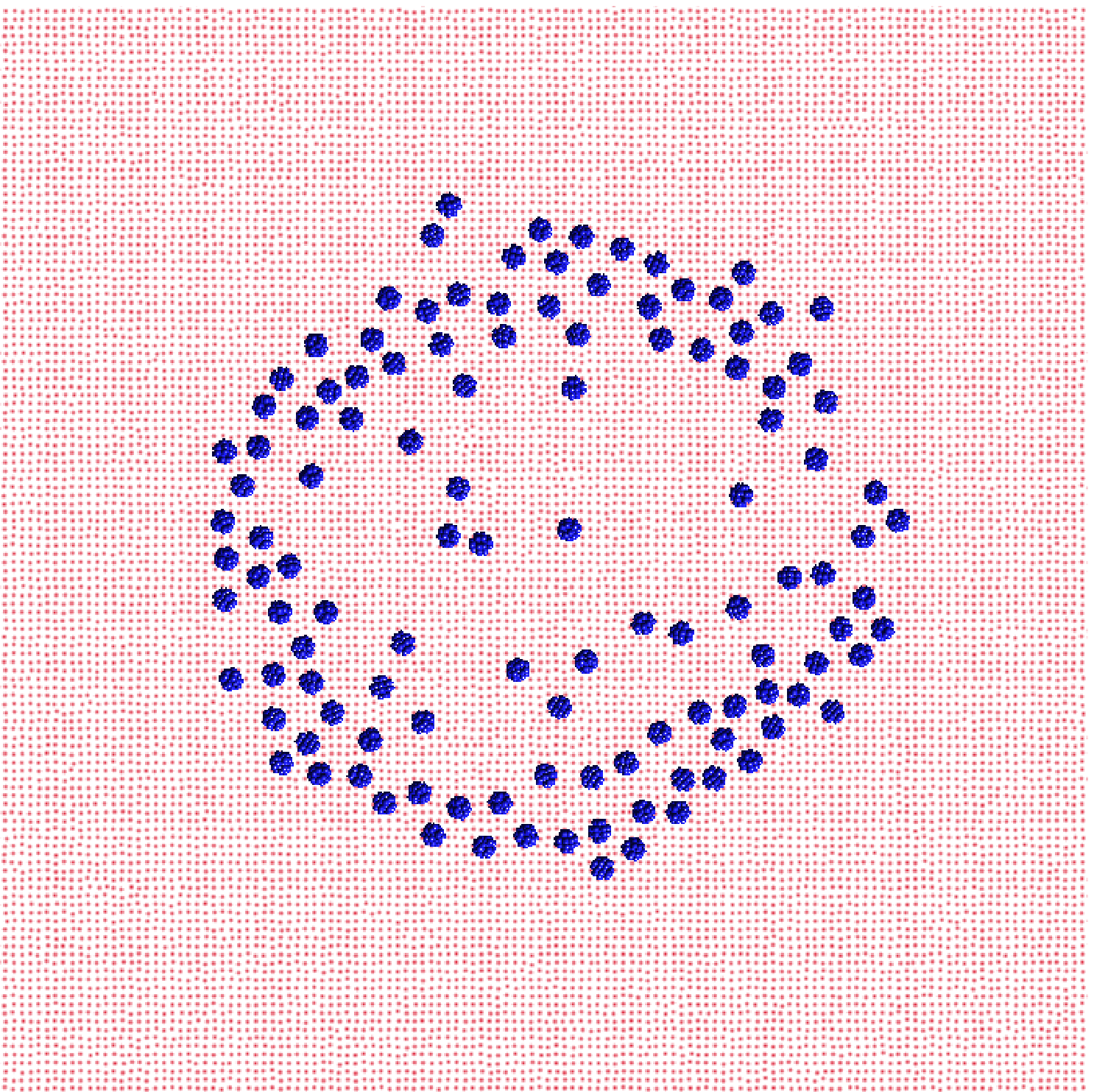}
  \includegraphics[width=0.30\textwidth, height=0.30\textwidth]{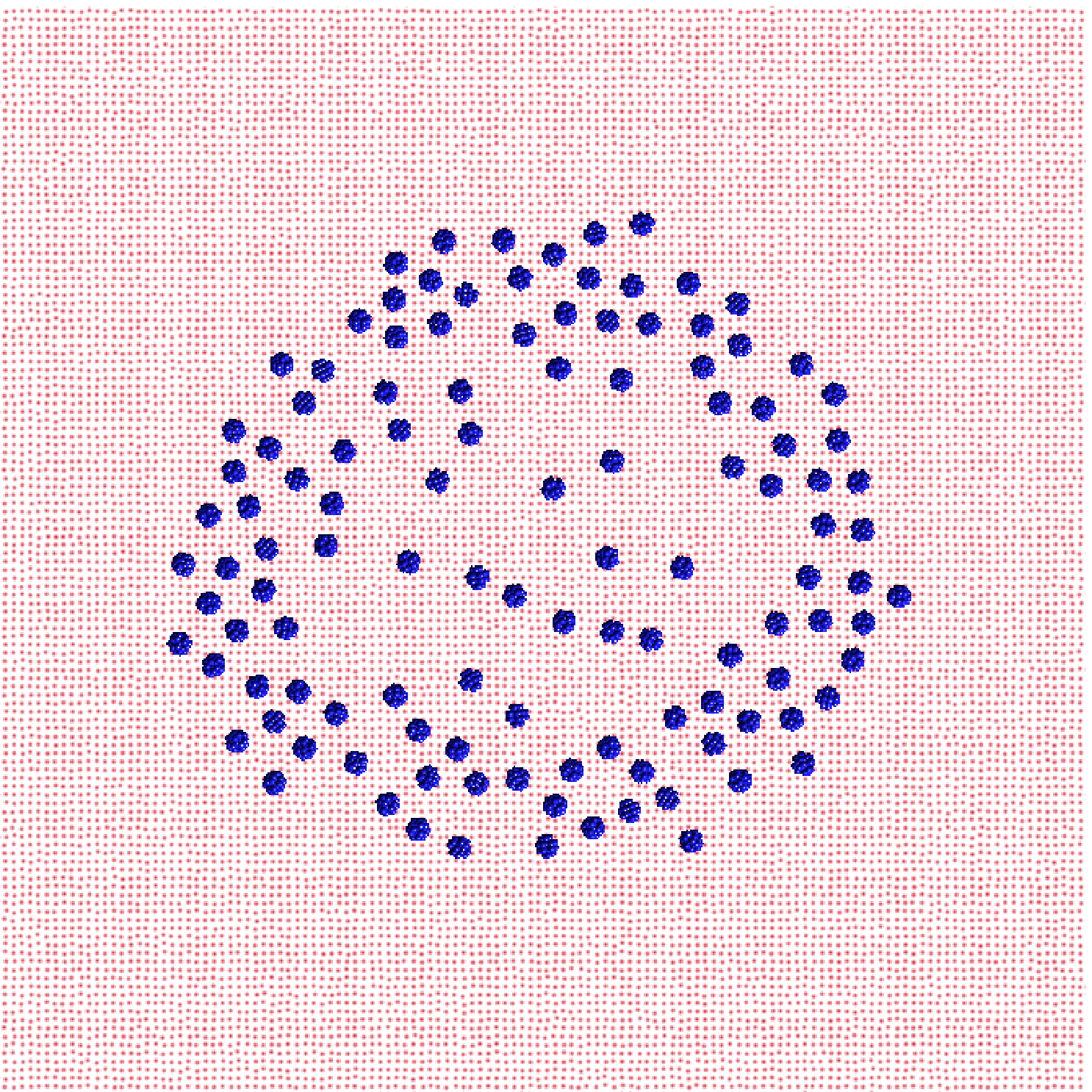} 
  \caption{Effects of adding a charge to a plain (non-Janus) particle.
  Top to bottom: charge 0, 4 and 8. \label{fig:plain}}
\end{figure} 

Additional experiments investigated the effect of particle surface charge on
the deposition behavior \cite{Rodner,Guzman}. The presence of sulfate groups 
on the 
surface of the particles results in an overall negative charge, which can be 
screened by the addition of salt.
Sulfated polystyrene particles in distilled water 
form hexagonal close-packed, highly ordered layers at the rim,
whereas formation of randomly packed particle layers is observed in
10 mM aqueous NaCl solution. These results motivated us to 
study the effect of charge
on the deposit structure of uniform particles. In Fig.~\ref{fig:plain}, we
show the deposits that result when the charge on a symmetric particle 
increases from 0 to 4 and then to 8.  In the simulations, a charge of the
appropriate magnitude is placed at the center of the particle, and a Coulomb
interaction is applied. The neutral case
resembles the random packing observed in the 10 mM case. The charge 8
situation models the case where the particles carry a charge, {\sl i.e.},
deposition in deionized water. It is apparent from the bottom panel of 
Fig.~\ref{fig:plain} that the particles tend
toward assembling at the rim and a regular packing \cite{Koh}, but that the 
drop volume is not sufficient to enable long enough evaporation times to 
achieve close packing at the rim.

%%%%%%%%%%%%%%%%%%%%%%%%%%%%%%%%%%%%%%%%%%%%%%%%%%%%%%%%%%%%%%%%%%%%%%%%%%%%%%%%
%                                                                              %
%                     Section III: Discussion and Conclusion                   %
%                                                                              %
%%%%%%%%%%%%%%%%%%%%%%%%%%%%%%%%%%%%%%%%%%%%%%%%%%%%%%%%%%%%%%%%%%%%%%%%%%%%%%%%

We have shown that straightforward, medium-scale MD simulations can easily 
capture most of the salient features in the evaporation of particle-laden
droplets. Aside from demonstrating that nano-scale and micron-scale systems
behave in a similar way with regard to the behavior of the particles,
we were able to measure continuum fields such as velocity within the droplet,
along with the profile of the evaporative flux, which drives the process.
Using standard methods, we have also measured (but not reported here) the
density, temperature, concentration and stress fields within the droplet.
In addition, we were able to show the existence of a minimum drop size 
for rim deposition, and verify the importance of adequate thermal coupling 
between liquid and solid.  Furthermore, the simulations give
the connection between particle interactions and deposit structure.
The significance of these results is that simple simulations provide a 
viable method for both testing the theoretical underpinnings of the process
and for predicting the nature of the outcome -- the structure of the
resulting particle deposit. In this paper we have focused on the most
important aspect of the continuum flow, the velocity field and the 
evaporative flux, but any other quantity which can be determined from atomic
variables is equally accessible.

\begin{acknowledgments}
This work was supported in part by NSF-CBET (CAREER) 0644789.  
\end{acknowledgments}

%\newpage

\end{document}